# What social media told us in the time of COVID-19: a scoping review

*Shu-Feng Tsao, Helen Chen, Therese Tisseverasinghe, Yang Yang, Lianghua Li, Zahid A Butt*

With the onset of the COVID-19 pandemic, social media has rapidly become a crucial communication tool for information generation, dissemination, and consumption. In this scoping review, we selected and examined peer-reviewed empirical studies relating to COVID-19 and social media during the first outbreak from November, 2019, to November, 2020. From an analysis of 81 studies, we identified five overarching public health themes concerning the role of online social media platforms and COVID-19. These themes focused on: surveying public attitudes, identifying infodemics, assessing mental health, detecting or predicting COVID-19 cases, analysing government responses to the pandemic, and evaluating quality of health information in prevention education videos. Furthermore, our Review emphasises the paucity of studies on the application of machine learning on data from COVID-19-related social media and a scarcity of studies documenting real-time surveillance that was developed with data from social media on COVID-19. For COVID-19, social media can have a crucial role in disseminating health information and tackling infodemics and misinformation.

**School of Public Health and Health Systems** (S-F Tsao MSc, Prof H Chen PhD, Y Yang MSc, Z A Butt PhD) **and Faculty of Science** (L Li BSc), **University of Waterloo, Waterloo, ON, Canada; Seneca Libraries, Seneca College, King City, ON, Canada** (T Tisseverasinghe MLIS)

Correspondence to:
Dr Zahid A Butt, School of Public Health and Health Systems, University of Waterloo, Waterloo, ON N2L 3G1, Canada
zahid.butt@uwaterloo.ca

## Introduction

Severe acute respiratory syndrome coronavirus 2 (SARS-CoV-2), and the resulting COVID-19, is a substantial international public health issue. As of Jan 18, 2021, an estimated 95 million people worldwide had been infected with the virus, with about 2 million deaths.[1] As a consequence of the pandemic, social media is becoming the platform of choice for public opinions, perceptions, and attitudes towards various events or public health policies regarding COVID-19.[2] Social media has become a pivotal communication tool for governments, organisations, and universities to disseminate crucial information to the public. Numerous studies have already used social media data to help to identify and detect outbreaks of infectious diseases and to interpret public attitudes, behaviours, and perceptions.[3–6] Social media, particularly Twitter, can be used to explore multiple facets of public health research. A systematic review identified six categories of Twitter use for health research, namely content analysis, surveillance, engagement, recruitment, as part of an intervention, and network analysis of Twitter users.[5] However, this review included only broader research terms, such as health, medicine, or disease, by use of Twitter data and did not focus on specific disease topics, such as COVID-19. Another article analysed tweets on COVID-19 and identified 12 topics that were categorised into four main themes: the origin, source, effects on individuals and countries, and methods of decreasing the spread of SARS-CoV-2.[7] In this study, data were not available for tweets that were related to COVID-19 before February, 2020, thereby missing the initial part of the epidemic, and the data for tweets were limited to between Feb 2 and March 15, 2020.

Social media can also be effectively used to communicate health information to the general public during a pandemic. Emerging infectious diseases, such as COVID-19, almost always result in increased usage and consumption of media of all forms by the general public for information.[8] Therefore, social media has a crucial role in people's perception of disease exposure, resultant decision making, and risk behaviours.[9,10] As information on social media is generated by users, such information can be subjective or inaccurate, and frequently includes misinformation and conspiracy theories.[11] Hence, it is imperative that accurate and timely information is disseminated to the general public about emerging threats, such as SARS-CoV-2. A systematic review explored the major approaches that were used in published research on social media and emerging infectious diseases.[12] The review identified three major approaches: assessment of the public's interest in, and responses to, emerging infectious diseases; examination of organisations' use of social media in communicating emerging infectious diseases; and evaluation of the accuracy of medical information that is related to emerging infectious diseases on social media. However, this review did not focus on studies that used social media data to track and predict outbreaks of emerging infectious diseases.

Analysing and disseminating information from peer-reviewed, published research can guide policy makers and public health agencies to design interventions for accurate and timely knowledge translation to the general public. Therefore, keeping in view the limitations of existing research that we have previously mentioned, we did a scoping review with the aim of understanding the roles that social media has had since the beginning of the COVID-19 crisis. We investigated public attitudes and perceptions towards COVID-19 on social media, information about COVID-19 on social media, use of social media for prediction and detection of COVID-19, the effects of COVID-19 on mental health, and government responses to COVID-19 on social media. Our objective was to identify and analyse studies on social media that were related to COVID-19 and focused on five themes: infodemics, public attitudes, mental health, detection or prediction of COVID-19 cases, government responses to the pandemic, and quality of health information in videos.

**Panel:** Variations of keywords and indexed terms that were used in the literature search

**Keywords**

*"COVID-19"*
- "Betacoronavirus", "severe acute respiratory syndrome", "covid 2019", "COVID-19", "corona-19", "n-cov", "novel coronavirus", "sars-cov", or "wuhan 2019"

*"Social media"*
- "Twitter", "tweet", "retweet", "facebook", "weibo", "sina", "youtube", "webcast", "user comment", "online post", "online discussion", "social network", "social media", "online community", or "mobile app"

**Indexed terms**

*"COVID-19"*
- "Betacoronavirus", "coronavirus infections", "severe acute respiratory syndrome", "coronavirus disease 2019", "COVID-19", or "covid-19"

*"Social media"*
- "Social networking", "social media", "mobile applications", "blogging", "social networking (online)", "online social network", "webcast", "mobile applications", "mobile computing", or "social network"

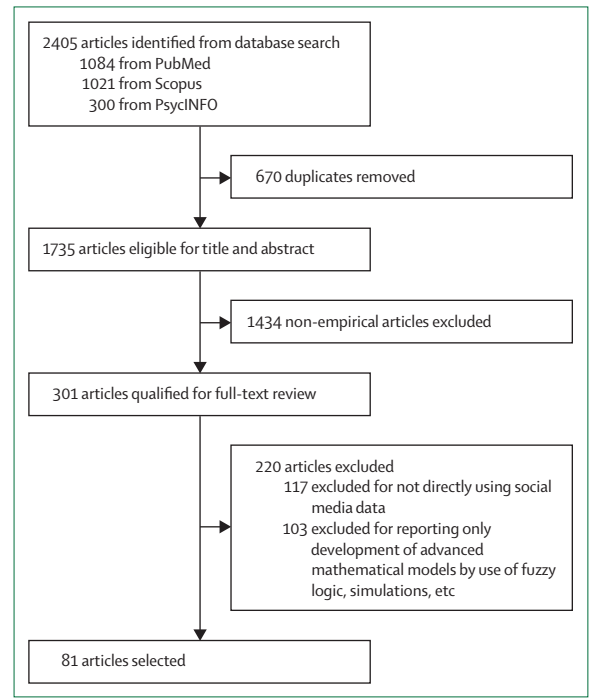

*Figure 1:* Preferred Reporting Items for Systematic Reviews and Meta-Analyses flow chart of article extraction from the literature search

## Methods

### Overview

Studies exploring the use of social media relating to COVID-19 were reviewed by use of the scoping review methods of Arksey and O'Malley[13] and Levac and colleagues.[14] We followed the five-step scoping review protocol and the Preferred Reporting Items for Systematic Reviews and Meta-Analyses extension for scoping reviews.

### Data Sources

Exploratory searches were done on COVID-19 Open Research Dataset Challenge and Google Scholar in April, 2020. These searches helped to define the Review scope, develop the research questions, and determine eligibility criteria. After such activity, MEDLINE and PubMed, Scopus, and PsycINFO were selected for this Review because they include peer-reviewed literature in the fields of medicine, behavioural sciences, psychology, health-care systems, and clinical sciences. Variations of the key search terms can be found in the panel. Since the start of the current pandemic, COVID-19 articles were reviewed and published at an unprecedently rapid rate, with numerous publications that were available ahead of print referred to as preprints or articles in press. In this Review, we consider peer-reviewed preprints to be equivalent to published peer-reviewed articles, and relevant articles were screened accordingly.

### Screening procedure

Mainly, the primary reviewer (S-FT) screened title and abstract for each article to decide whether an article met the inclusion criteria. If the criteria were confirmed, then the article was included; otherwise, it was excluded. Paragraphs in articles were assigned a code representing one of the five themes (eg, I for infodemic), then a code was assigned to the article on the basis of the majority of paragraph codes. Next, quotes were sorted under each code, applying Ose's method.[15] Braun and Clark's thematic analysis method was used and involved searching for the text that matched the identified predictors (ie, codes) from the quantitative analysis and discovering emergent codes that were relevant to either the study objective or identified in the relevant literature review.[16] Finally, we categorised the codes into main themes. These codes and themes were compared and clarified by S-FT, ZAB, and YY to draw conclusions around the main themes. S-FT is fluent in English and Mandarin. The secondary reviewer (ZAB) is fluent in English, and the tertiary reviewer and domain expert (YY and HC) are both fluent in English and Mandarin. Any discrepancies among reviewers were discussed with the research team to reach consensus.

### Results

With the application of appropriate search filters, a total of 2405 articles were retrieved from the identified databases: PubMed (1084 articles), Scopus (1021 articles), and PsycINFO (300 articles). Among these, 670 duplicates were excluded. Of the remaining 1735 articles, 1434 were deemed to be non-empirical, such as comments, editorial essays, letters, opinions, and reviews. These exclusions left 301 articles for a full-text review on the basis of the screening results of titles and abstracts. After the full-text review, 81 articles were included in this scoping review (figure 1).

|  | Publication month | Origin | Social media | Study population and sample size | Methods | Key findings |
|---|---|---|---|---|---|---|
| **Detection or prediction of COVID-19 cases** | | | | | | |
| Li et al[17] | March | China | Google Trends, Baidu Search Index, and Sina Weibo Index | Keywords of coronavirus and pneumonia were searched, and trend data was collected from Google Trends, Baidu Search Index, and Sina Weibo Index from Jan 2 to Feb 20, 2020 | Lag correlation | Lag correlations showed a maximum correlation between trend data and the number of diagnoses at 8–12 days before for laboratory-confirmed cases and 6–8 days before for suspected cases |
| Liu et al[18] | August | China | Sina Weibo | Sina Weibo messages between Jan 20 and Feb 15, 2020; 599 participants | Gathered data via Sina Weibo, then followed up with telephone call; statistical analysis taken with Fisher exact test; rates of death calculated with Kaplan-Meier method; multivariate Cox regression used to establish risk factors for mortality | Older age (ie, >69 years), diffuse pneumonia, and hypoxaemia are factors that can help clinicians to identify patients with COVID-19 who have poor prognosis; aggregated data from social media can also be comprehensive, immediate, and informative in disease prognosis |
| O'Leary and Storey[19] | September | USA | Google Trends, Wikipedia, and Twitter | Google Trends searches for coronavirus and COVID-19 between Jan 21 and April 5, 2020; Wikipedia page views for coronavirus and COVID-19 between Jan 12 and April 5, 2020; number of Twitter original tweets between Jan 27 and April 5, 2020; numbers of COVID-19 cases and deaths in the USA[20] | Regression analysis | To model the number of cases, the current Wikipedia page views, tweets from 1 week before, and Google Trends searches from 2 weeks before were used; to model of the number of deaths, each variable was taken from 1 week earlier than for cases |
| Peng et al[21] | June | China | Sina Weibo | 1200 records | Spatiotemporal distribution of COVID-19 cases in the main urban area of Wuhan, China; kernel density analysis; ordinary least square regression | Older people (ie, >60 years) are at high risk of severe symptoms and have high prevalence in the COVID-19 outbreak, and they account for >50% of the total number of Sina Weibo help seekers; early transmission of COVID-19 in Wuhan, China, could be divided into three phrases: scattered infection, community spread, and full-scale outbreak |
| Qin et al[22] | March | China | Baidu Search Index | Social media search index for dry cough, fever, chest distress, coronavirus, and pneumonia from Dec 31, 2019, to Feb 9, 2020; data for new suspected cases of COVID-19 from Jan 20 to Feb 9, 2020 | Subset selection; forward selection; lasso regression; ridge regression; elastic net | Case numbers of new suspected COVID-19 correlated significantly with the lagged series of social media search index; social media search index could detect new suspected COVID-19 cases 6–9 days earlier than could laboratories |
| Zhu et al[23] | April | China | Sina Weibo | 1101 Sina Weibo posts related to COVID-19 from Dec 31, 2019, to Feb 12, 2020 | Descriptive statistics: numbers and percentage; time series analysis | Attention to COVID-19 was low until China openly admitted human-to-human transmission on Jan 20, 2020; attention quickly increased and remained high over time |
| **Government responses** | | | | | | |
| Basch et al[24] | April | USA | YouTube | 100 most widely viewed videos uploaded in January, 2020 | Descriptive analysis: frequency, percentage, mean, and standard deviation | Percentage of each of the seven key prevention behaviours that are listed on the US Centers for Disease Control and Prevention website that were covered in the 100 videos varied from 0% (eg, use a face mask for protection if you are caring for the ill) to 31% (avoid close contact with people who are sick); overall, videos that covered at least one prevention behaviour accounted for less than one-third of the 100 videos |
| Basch et al[25] | April | USA | YouTube | 100 most widely viewed YouTube videos as of Jan 31, 2020, and March 20, 2020, with keyword of coronavirus in English, with English subtitles, or in Spanish | Descriptive analysis: frequency, percentage, mean, and standard deviation | <50% of videos in either sample covered any of the prevention behaviours that are recommended by the US Centers for Disease Control and Prevention |
| Khatri et al[26] | March | Singapore | Youtube | 150 videos collected on Feb 1–2, 2020, with keywords of 2019 novel coronavirus (50 videos), and Wuhan virus in English (50 videos) and Mandarin (50 videos) | Descriptive analysis: percentage and mean; DISCERN score; Medical Information and Content Index score | Mean DISCERN score for reliability was 3·12 of 5·00 for English and 3·25 of 5·00 for Mandarin videos; mean cumulative Medical Information and Content Index score of useful videos was 6·71 of 25·00 for English and 6·28 of 25·00 for Mandarin |

(Table continues on next page)

|  | Publication month | Origin | Social media | Study population and sample size | Methods | Key findings |
|---|---|---|---|---|---|---|
| (Continued from previous page) | | | | | | |
| Li et al[27] | March | China | Sina Weibo | 36 746 Sina Weibo data from Dec 30, 2019, to Feb 1, 2020; a random sample of 3000 Sina Weibo posts as training dataset | Linear regression; support vector machine; Naive Bayes; natural language processing | Classified the information related to COVID-19 into seven types of situational information and their predictors |
| Merkley et al[28] | April | Canada | Twitter and Google Trends | 33 142 tweets from 292 social media accounts of federal members of parliament from Jan 1 to March 28, 2020; 87 Google search trends for the search term coronavirus in the first half (ie, days 1–14) and second half (ie, days 15–31) of March, 2020; a survey of 2499 Canadian citizens ≥18 years from April 2 to April 6, 2020 | Linear regression | No members of parliament from any party downplaying the pandemic; no association between Conservative Party vote share and Google search interest in the coronavirus |
| Rufai and Bunce[29] | April | USA | Twitter | 203 viral tweets from G7 world leaders from Nov 17, 2019, to March 17, 2020 with keywords COVID-19 or coronavirus and a minimum of 500 likes | Qualitative design; content analysis | 166 of 203 of tweets were informative; 9·4% (19) were morale-boosting; 6·9% (14) were political |
| Sutton et al[30] | September | USA | Twitter | 690 accounts representing public health, emergency management, and elected officials and 149 335 tweets | $\chi^2$ analyses; negative binomial regression modelling | Systematic changes were made in message strategies over time and identified key features that affect message passing, both positively and negatively; results have the potential to aid in message design strategies as the pandemic continues, or in similar future events |
| Wang et al[31] | September | USA | Twitter | 13 598 tweets related to COVID-19 from Jan 1 to April 27, 2020 | Temporal analysis and networking analysis | 16 categories of message types were manually annotated; inconsistencies and incongruencies were identified in four critical topics (ie, wearing masks, assessment of risks, stay at home order, and disinfectant and sanitiser); network analysis showed increased communication coordination over time |
| Infodemics | | | | | | |
| Ahmed et al[32] | October | UK | Twitter | 22 785 tweets and 11 333 Twitter users with #FilmYourHospital from April 13 to April 20, 2020 | Social network analysis; user analysis | The most important drivers of the #FilmYourHospital conspiracy theory are ordinary citizens; YouTube was the information source most linked to by users; the most retweeted post belonged to a verified Twitter user |
| Ahmed et al[33] | May | UK | Twitter | A subsample of 233 tweets from 10 140 tweets collected from 19:44 h UTC on Friday, March 27, 2020, to 10:38 h UTC on Saturday, April 4, 2020 were used for content analysis | Descriptive statistics: numbers, percentage; social network analysis; content analysis | 34·8% (81 of 233) of tweets linked 5G and COVID-19; 32·2% (75) of tweets denounced the conspiracy theory |
| Brennen et al[34] | October | UK | Digital visual media | 96 samples of visuals from January to March, 2020 | Qualitative coding | Organised all findings into six trends: authoritative agency, virulence, medical efficacy, intolerance, prophecy, satire; a small number of manipulated visuals, all were produced by use of simple tools; no examples of so-called deepfakes (ie, techniques that are used to make synthetic videos that closely resemble real videos) or other techniques that were based on artificial intelligence |
| Bruns et al[35] | August | Australia | Facebook | 89 664 distinct Facebook posts from Jan 1 to April 12, 2020 | Time series; network analysis | Substantially increased number of posts about 5G rumours on Facebook after March 19, 2020; network analysis showed that coalitions of various groups were brought together by conspiracy theories about COVID-19 and 5G technology |
| Galhardi et al[36] | October | Brazil | WhatsApp, Instagram, and Facebook | Fake news collected from March 17 to April 10, 2020, on the basis of data from the Eu Fiscalizo app (version 5.0.5) | Quantitative content analysis | WhatsApp is the main channel for sharing fake news, followed by Instagram and Facebook |
| Gallotti et al[37] | October | Italy | Twitter | >100 million Tweets | Developed an Infodemic Risk Index | Before the rise of COVID-19 cases, entire countries had measurable waves of potentially unreliable information, posing a serious threat to public health |

(Table continues on next page)

|  | Publication month | Origin | Social media | Study population and sample size | Methods | Key findings |
|---|---|---|---|---|---|---|
| (Continued from previous page) | | | | | | |
| Islam et al[38] | October | Bangladesh | Fact-checking agency websites, Facebook, Twitter, and websites for television networks and newspapers | 2311 infodemic reports related to COVID-19 between Dec 31, 2019, and April 5, 2020 | Descriptive analysis; spatial distribution analysis | Misinformation that is fuelled by rumours, stigma, and conspiracy theories can have potentially severe implications on public health if prioritised over scientific guidelines; governments and other agencies should understand the patterns of rumours, stigma, and conspiracy theories that are related to COVID-19 and circulating globally so that they can develop appropriate messages for risk communication |
| Kouzy et al[39] | March | Lebanon | Twitter | 673 English tweets collected on Feb 27, 2020; 617 tweets after exclusion of tweets that were humorous or not serious | Descriptive statistics; bar chart; $\chi^2$ statistic to calculate p value (2-sided; p=0·05 significance threshold) for the association between account or tweet characteristics and the presence of misinformation or unverifiable information about COVID-19 | 153 (24·8%) of 617 tweets had misinformation; 107 (17·3%) had unverifiable information; misinformation rate higher in informal individual or group accounts than in formal individual or group accounts (33·8% [123 of 364] vs 15·0% [30 of 200], p<0·0010) |
| Moscadelli et al[40] | August | Italy | Fake news and corresponding verified news that was circulated in Italy | 2102 articles between Dec 31, 2019, and April 30, 2020 | Social media trend analysis by use of BuzzSumo | Links containing fake news were shared 2 352 585 times, accounting for 23·1% (2 352 585 of 10 184 351) of total shares of all reviewed articles |
| Pulido et al[41] | April | Spain | Twitter | 942 valid tweets between Feb 6 and Feb 7, 2020 | Communicative content analysis | Misinformation was tweeted more but retweeted less than tweets based on scientific evidence; tweets based on scientific evidence had more engagement than misinformation |
| Rovetta and Bhagavathula[42] | August | Italy | Google Trends and Instagram | 2 million Google Trends queries and Instagram hashtags from Feb 20 to May 6, 2020 | Classification of infodemic monikers (ie, a term, query, hashtag, or phrase that generates or feeds fake news, misinterpretations, or discrimination); computed the mean peak volume with a 95% CI | Globally, growing interest exists in COVID-19, and numerous infodemic monikers continue to circulate on the internet |
| Uyheng and Carley[43] | October | USA and Philippines | Twitter | 12·0 million tweets from 1·6 million users from the USA and 15·0 million tweets from 1·0 million users from the Philippines between March 5 and March 19, 2020 | Hate speech score assigned to each tweet by use of machine learning algorithm; bot scores were assigned to each user via BotHunter algorithm; social media analysis via ORA software; network analysis via centrality analysis; cluster analysis via Leiden algorithm | Analysis showed idiosyncratic relationships between bots and hate speech across datasets, emphasising different network dynamics of racially charged toxicity in the USA and political conflicts in the Philippines; bot activity is linked to hate in both countries, especially in communities that are dense and isolated from others |
| Mental health | | | | | | |
| Gao et al[44] | April | China | Sina Weibo | Online survey on Wenjuanxing platform from Jan 31 to Feb 2, 2020; with 4872 Chinese citizens aged ≥18 years from 31 provinces and autonomous regions in China | Multivariable logistic regression | Social media exposure was frequently positively associated with high odds of anxiety (odds ratio 1·72, 95% CI 1·31–2·26) and combination of depression and anxiety (odds ratio 1·91, 95% CI 1·52–2·41) |
| Li et al[45] | March | China | Sina Weibo | Sina Weibo posts from 17 865 active Sina Weibo users between Jan 13 and Jan 26, 2020 | Sentiment analysis; paired sample t-test | Negative emotions and sensitivity to social risks increased; scores of positive emotions and life satisfaction decreased after outbreak declaration |

(Table continues on next page)

The table summarises the 81 articles that were selected on COVID-19 and social media. All articles were written in English. Data from Twitter (45 articles) and Sina Weibo (16 articles) were undoubtedly the most frequently studied. To categorise these chosen articles, we adopted a novel framework called Social Media and Public Health Epidemic and Response (SPHERE) and developed a modified version of SPHERE framework to organise the themes for our scoping review (figure 2).[98] Themes were identified through reviewers' consensus based on our

| | Publication month | Origin | Social media | Study population and sample size | Methods | Key findings |
|---|---|---|---|---|---|---|
| (Continued from previous page) | | | | | | |
| **Prevention education in videos** | | | | | | |
| Hakimi and Armstrong[46] | September | USA | YouTube | 49 of the first 100 videos on YouTube with the most views that were identified by the search term DIY hand sanitiser; 51 videos were excluded because they were not in English or not related to the search term | Codified video content; assessed by use of Cohen's κ; descriptive statistics calculated; assessed by $\chi^2$ test with 2-sided p value <0·05 as the threshold for significance | Most videos did not describe labelling storage containers, 69% (34 of 49) of videos encouraged the use of oils or perfumes to enhance hand sanitiser scent, and 2% (1) of videos promoted the use of colouring agents to be more attractive for use among children specifically; significantly increased mean number of daily calls to poison control centres regarding unsafe paediatric exposure to hand sanitiser since the first confirmed patient with COVID-19 in the USA (p<0·0010); significantly increased mean number of daily calls in March, 2020, compared with the previous 2 years (p<0·0010) |
| Hernández-García and Giménez-Júlvez[47] | June | Spain | YouTube | 129 videos in Spanish with the terms prevencion coronavirus and prevencion COVID19 | Univariate analysis; multiple logistic regression model | Information from YouTube in Spanish on basic measures to prevent COVID-19 is usually not complete and differs according to the type of authorship (ie, mass media, health professionals, individual users, or others) |
| Moon and Lee[48] | August | South Korea | YouTube | 105 most viewed YouTube videos from Jan 1 to April 30, 2020 | Modified DISCERN index; *Journal of the American Medical Association* Score benchmark criteria; Global Quality Score; Title–Content Consistency Index; Medical Information and Content Index | 37·14% (39 of 105) of videos contained misleading information; independent user-generated videos showed the highest proportion of misleading information at 68·09% (32 of 47); misleading videos had more likes, fewer comments, and longer running times than did useful videos; transmission and precautionary measures were the most frequently covered content |
| Ozdede and Peker[49] | July–August | Turkey | YouTube | The top 116 English language videos with at least 300 views | Precision indices and total video information and quality index scores were calculated | High number of views on dentistry YouTube videos related to COVID-19; quality and usefulness of these videos are moderate |
| Yüce et al[50] | July | Turkey | YouTube | 55 English videos about COVID-19 control procedures for dental practices collected on March 31, 2020, between 9:00 h and 18:00 h | Modified DISCERN instrument; descriptive statistics | Only two (3·6%) of 55 videos were good quality, whereas 24 (43·6%) videos were poor quality |
| **Public attitudes** | | | | | | |
| Abd-Alrazaq et al[7] | April | Qatar | Twitter | 2·8 million English tweets (167 073 unique tweets from 160 829 unique users) from Feb 2 to March 15, 2020 | Word frequencies of single (ie, unigrams) and double words (ie, bigrams); sentiment analysis; mean number of retweets, likes, and followers for each topic; interaction rate per topic; LDA for topic modelling | Identified 12 topics and grouped into four themes; average sentiment positive for ten topics and negative for two topics |
| Al-Rawi et al[51] | November | Canada | Twitter | Over 50 million tweets referencing #Covid-19 and #Covid19 for more than 2 months in early 2020 | Mixed method: analysed emoji use by each gender category; the top 600 emojis were manually classified on the basis of their sentiment | Identified five major themes in the analysis: morbidity fears, health concerns, employment and financial issues, praise for front-line workers, and unique gendered emoji use; most emojis are extremely positive across genders, but discussions by women and gender minorities are more negative than by men; when discussing particular topics (eg, financial and employment matters, gratitude, and health care), there are many differences; use of several unique gender emojis to express specific issues (eg, coffin, skull, and siren emojis were used more often by men than by other genders when discussing fears and morbidity, whereas the use of the folded hands emoji as a thankful gesture for front-line workers was found more often in discussions by women than by other genders and the bank emoji was noted only in women's discussions) |

modified SPHERE framework. We identified six themes: infodemics, public attitudes, mental health, detecting or predicting COVID-19 cases, government responses, and quality of health information in prevention education videos.

### Social media as contagion and vector

According to WHO, the term infodemic, a combination of information and epidemic, refers to a fast and widespread dissemination of both accurate and inaccurate information about an epidemic, such as

| | Publication month | Origin | Social media | Study population and sample size | Methods | Key findings |
|---|---|---|---|---|---|---|
| (Continued from previous page) | | | | | | |
| Arpaci et al[52] | July | Turkey | Twitter | 43 million tweets between March 22 and March 30, 2020 | Evolutionary clustering analysis | Unigram terms appear more frequently than bigram and trigram (ie, triple words) terms; during the epidemic, many tweets about COVID-19 were distributed and attracted widespread public attention; high-frequency words (eg, death, test, spread, and lockdown) indicated that people were afraid of being infected and people who were infected were afraid of death; people agreed to stay at home due to fear of spread and called for physical distancing since they became aware of COVID-19 |
| Barrett et al[53] | August | USA | Twitter | 188 tweets about Governor Dan Patrick's statement on March 23, 2020, about generational self-sacrifice. | Thematic analysis | 90% (169 of 188) of tweets opposed calculated ageism, whereas only 5% (9) supported it and 5% (10) conveyed no position; opposition centred on moral critiques, political–economic critiques, assertions of the worth of older adults (eg, >60 years), and public health arguments; support centred on individual responsibility and patriotism |
| Boon-Itt and Skunkan[54] | November | Thailand | Twitter | 107 990 English tweets related to COVID-19 between Dec 13, 2019, and March 9, 2020 | Sentiment analysis; topic modelling by use of LDA | Sentiment analysis showed a predominantly negative feeling towards the COVID-19 pandemic; topic modelling revealed three themes relating to COVID-19 and the outbreak: the COVID-19 pandemic emergency, how to control COVID-19, and reports on COVID-19 |
| Budhwani and Sun[55] | May | USA | Twitter | 16 535 tweets about Chinese virus or China virus between March 9 and March 15, 2020, 177 327 tweets between March 19 and March 25, 2020 | Descriptive analysis; spatial analysis | Nearly 10 times increase at the national level; all 50 states had an increase in the number of tweets exclusively mentioning Chinese virus or China virus instead of coronavirus disease, COVID-19, or coronavirus; mean 0·38 tweets referencing Chinese virus or China virus were posted per 10 000 people at the state level in the preperiod (ie, March 9–15, 2020), and 4·08 of these stigmatising tweets were posted in the postperiod (ie, March 19–25, 2020), also indicating a 10 times increase |
| Chang et al[56] | November | Taiwan | 10 news websites, 11 discussion forums, 1 social network, 2 principal media sharing networks | 1·07 million Chinese texts from Dec 30, 2019, to March 31, 2020 | Deductive analysis | Online news promoted negativity and drove emotional social posts; stigmatising language that was linked to the COVID-19 pandemic showed an absence of civic responsibility that encouraged bias, hostility, and discrimination |
| Chehal et al[57] | July | India | Twitter | 29 554 tweets during the second lockdown (ie, April 15–May 3, 2020); 47 672 tweets during the third lockdown (May 4–17, 2020) | Sentiment analysis by use of the National Research Council of Canada Emotion Lexicon | A positive approach in the second lockdown but a negative approach in the third lockdown |
| Chen et al[58] | September | China | Sina Weibo | 1411 posts pertinent to COVID-19 taken from Healthy China, an official Sina Weibo account of the National Health Commission of China, from Jan 14 to March 5, 2020 | Descriptive analysis; hypothesis testing | Media richness (ie, potential information load, where low richness is only text and high richness is not only text) negatively predicted citizen participation via government social media, but dialogic loop (ie, stimulation of public dialogue, provision of the dialogue channel, and response to public feedback in a timely manner) facilitated engagement |
| Damiano and Allen Catellier[59] | August | USA | Twitter | 600 English tweets from the USA were selected: 300 from February, 2020, and 300 from March, 2020 | Frequencies; $\chi^2$ statistics | Neutral sentiment; tweets about COVID-19 risks and emotional outrage accounted for <50% (135 of 600); few tweets were related to blame |
| Darling-Hammond et al[60] | September | USA | Twitter | 339 063 tweets from non-Asian respondents of the Project Implicit Asian Implicit Association Test from 2007–20 and were broken into two datasets: the first dataset was from Jan 1, 2007, to Feb 10, 2020; the second data set was from Feb 11 to March 31, 2020 | Local polynomial regression; interrupted time-series analyses | Implicit Americanness Bias steadily decreased from 2007 to 2020; when media entities began using stigmatising terms, such as Chinese virus, starting from March 8, 2020, Implicit Americanness Bias began to increase; such bias was more pronounced among conservative individuals than among non-conservative individuals |

(Table continues on next page)

| | Publication month | Origin | Social media | Study population and sample size | Methods | Key findings |
|---|---|---|---|---|---|---|
| (Continued from previous page) | | | | | | |
| Das and Dutta[61] | July | India | Twitter | 410 643 tweets with #IndiaLockdown and #IndiafightsCorona from March 22 to April 21, 2020 | National Research Council of Canada lexicon for corpus-level emotion mining; sentimentr from open source R software for sentiment analysis to create additional sentiment scores; LDA for topic models; Natural Language Toolkit to develop sentiment-based topic models | For the broad corpus-level analysis, the context of positiveness was substantially higher than were negative sentiments; however, positive sentiment trends were similar to negative sentiment trends in terms of topics covered when the analysis was done at individual tweet level; the results showed that the discussion of COVID-19 in India on Twitter contains slightly more positive sentiments than negative sentiments |
| De Santis et al[62] | July | Italy | Twitter | 1 044 645 tweets | A general purpose methodological framework, grounded on a biological metaphor and on a chain of NLP and graph analysis techniques | Energy evolution through time was monitored; daily hot topics were identified (eg, COVID-19, Walter Ricciardi's retweet of an anti-Trump tweet from Michael Moore, Gabriele Gravina's argument against suspension of Italian football, increased COVID-19 cases in Italy, high case numbers in Lombardy, Italy, and an interview of Matteo Salvini about COVID-19 topics by Massimo Giletti) |
| Dheeraj[63] | May–June | India | Reddit | 868 posts related to COVID-19 | Fetching the articles: Python Reddit Application Programming Interface Wrapper; data preprocessing: Reddit Application Programming Interface and Natural Language Toolkit library | Of 868 posts on Reddit that were related to COVID-19 articles, 50% (434) were neutral, 22% (191) were positive, and 28% (243) were negative |
| Essam and Abdo[64] | August | Egypt | Twitter | 1 920 593 tweets with corona, coronavirus, or COVID-19 keywords from Feb 1 to April 30, 2020 | Thematic analysis | The dominant themes that were closely related to coronavirus tweets included the outbreak of the pandemic, metaphysics responses, signs and symptoms in confirmed cases, and conspiracies; the psycholinguistic analysis showed that tweeters maintained high amounts of affective talk (ie, expression of feelings), which was loaded with negative emotions and sadness; Linguistic Inquiry and Word Count's psychological categories of religion and health dominated the Arabic tweets discussing the pandemic situation |
| Yin FL et al[65] | March | China | Sina Weibo | Sina Weibo posts from Dec 31, 2019, to Feb 7, 2020 | Multiple-information susceptible-discussing-immune model | Model reproduction ratio declined from 1·78 to 0·97, showing that the peak of posts had passed but the topic was still on social media afterwards with a decreased number of posts |
| Gozzi et al[66] | October | Italy, UK, USA, and Canada | News, YouTube, Reddit, and Wikipedia | 227 768 web-based news articles from Feb 7 to May 15, 2020; 13 448 YouTube videos from Feb 7 to May 15, 2020; 107 898 English user posts and 3 829 309 comments on Reddit from Feb 15 to May 15, 2020; 278 456 892 views of Wikipedia pages that were related to COVID-19 from Feb 7 to May 15, 2020 | Linear regression; topic modelling by use of LDA | Collective attention was mainly driven by media coverage rather than epidemic progression, rapidly became saturated, and decreased despite media coverage and COVID-19 incidence remaining high; Reddit users were generally more interested in health, data regarding the new disease, and interventions needed to halt the spreading with respect to media exposure than were users of other platforms |
| Green et al[67] | July | USA | Twitter | 19 803 tweets from Democrats and 11 084 tweets from Republicans between Jan 17 and March 31, 2020 | Random forest | Democrats discussed the crisis more frequently—emphasising public health and direct aid to US workers—whereas Republicans placed greater emphasis on national unit, China, and businesses |
| Han et al[68] | April | China | Sina Weibo | 1 413 297 Sina Weibo messages, including 105 330 texts with geographical location information, from 00:00 h on Jan 9, 2020, to 00:00 h on Feb 11, 2020 | Time series analysis; kernel density estimation; Spearman correlation; LDA model; random forest algorithm | Public response was sensitive to the epidemic and notable social events, especially in urban agglomerations |
| Jelodar et al[69] | June | China | Reddit | 563 079 English comments related to COVID-19 from Reddit between Jan 20 and March 19, 2020 | Topic modelling by use of LDA and probabilistic latent semantic analysis; sentiment classification by use of recurrent neural network | The results showed a novel application for NLP based on a long short term memory model to detect meaningful latent topics and sentiment–comment classification on issues related to COVID-19 on social media |

(Table continues on next page)

|  | Publication month | Origin | Social media | Study population and sample size | Methods | Key findings |
|---|---|---|---|---|---|---|
| (Continued from previous page) | | | | | | |
| Jimenez-Sotomayor et al[70] | April | Mexico | Twitter | A random sample of 351 of 18 128 tweets were analysed from March 12 to March 21, 2020 | Qualitative content classification | The most common types of tweets were personal opinions (31·9% [112 of 351]), followed by informative tweets (29·6% [104]), jokes or ridicule (14·2% [50]), and personal accounts (13·4% [47]); 72 of 351 tweets were most likely intended to ridicule or offend someone and 21·1% (74) had content implying that the life of older adults (ie, referred to in tweets as "elderly", "older", and "boomer") was less valuable than that of younger people or downplayed the relevance of COVID-19 |
| Kim[71] | August | South Korea | Twitter | 27 849 individual tweets about COVID-19 between Feb 10 and Feb 14, 2020 | Binary logistic regression; semantic network analysis | Social network size was a negative predictor of incivility |
| Kurten and Beullens[72] | August | Belgium | Twitter | 373 908 tweets and retweets from Feb 25 to March 30, 2020 | Time series; network bigrams; emotion lexicon; LDA | Notable COVID-19 events immediately increased the number of tweets; most topics focused on the need for EU collaboration to tackle the pandemic |
| Kwon et al[73] | October | USA | Twitter | 259 529 unique tweets containing the word coronavirus between Jan 23 and March 24, 2020 | Trending analysis; spatiotemporal analysis | Early facets of physical distancing appeared in Los Angeles (CA, USA), San Francisco (CA, USA), and Seattle (WA, USA); social disruptiveness tweets were most retweeted, and intervention implementation tweets were most favourited |
| Lai et al[74] | October | USA | Reddit | 522 comments from an Ask Me Anything session on COVID-19 on March 11, 2020, from 14:00 h to 16:00 h EST | Content analysis | The highest number of posts were about symptoms (27% [141 of 522]), followed by prevention (25% [131]); symptoms was the most common intended topic for further discussions (28% [94 of 337]) |
| Li et al[75] | April | China | Sina Weibo | 115 299 Sina Weibo posts from Dec 23, 2019, to Jan 30, 2020; 11 893 of them were collected from Dec 31, 2019, to Jan 20, 2020, for qualitative analysis; total daily cases of COVID-19 in Wuhan, China, were obtained from the Chinese National Health Commission | Linear regression model; qualitative content analysis | Positive correlation between the number of Sina Weibo posts and the number of reported cases, with ten COVID-19 cases per 40 posts; posts grouped into four themes |
| Li et al[76] | September | USA | Twitter | 155 353 unique English tweets related to COVID-19 that were posted from Dec 31, 2019, to March 13, 2020 | Content analysis | Peril of COVID-19 was mentioned the most often, followed by content about marks (ie, cues to identify members of a stigmatised group: flu-like symptoms, personal protective equipment, Asian origin, and health-care providers and essential workers), responsibility, and group labelling; information on conspiracy theories was more likely to be included in tweets about group labelling and responsibility than in tweets about COVID-19 peril |
| Lwin et al[77] | May | Singapore | Twitter | 20 325 929 tweets from 7 033 158 unique users from Jan 28 to April 9, 2020 | Sentiment analysis | Public emotions shifted strongly from fear to anger over the course of the pandemic, while sadness and joy also surfaced; anger shifted from xenophobia at the beginning of the pandemic to discourse around the stay-at-home notices; sadness was emphasised by the topics of losing friends and family members, whereas topics that were related to joy included words of gratitude and good health; emotion-driven collective issues around shared public distress experiences of the COVID-19 pandemic are developing and include large-scale social isolation and the loss of human lives |
| Ma et al[78] | July | China | WeChat | Top 200 accounts from Jan 21 to Jan 27, 2020 | Simple linear regression; multiple linear regression; content analysis | For non-medical institution accounts in the model, report and story types of articles had positive effects on whether users followed behaviours; for medical institution accounts, report and science types of articles had a positive effect |
| Medford et al[79] | June | USA | Twitter | 126 049 English tweets from 53 196 unique users with matching hashtags that were related to COVID-19 from Jan 14 to Jan 28, 2020 | Temporal analysis; sentiment analysis; topic modelling by use of LDA | The hourly number of tweets that were related to COVID-19 starkly increased from Jan 21, 2020, onwards; fear was the most common emotion and was expressed in 49·5% (62 424 of 126 049) of all tweets; the most common predominant topic was the economic and political effect |
| Mohamad[80] | June | Brunei | Twitter, Instagram, and TikTok | 30 individual profiles from Instagram, Twitter, and TikTok | Qualitative content analysis | Five narratives of local responses to physical distancing practices were apparent: fear, responsibility, annoyance, fun, and resistance |

(Table continues on next page)

| | Publication month | Origin | Social media | Study population and sample size | Methods | Key findings |
|---|---|---|---|---|---|---|
| (Continued from previous page) | | | | | | |
| Nguyen et al[81] | September | USA | Twitter | 3 377 295 US tweets that were related to race from November, 2019, to June, 2020 | Support vector machine was used for sentiment analysis | Proportion of negative tweets referencing Asians increased by 68·4%; proportion of negative tweets referencing other racial or ethnic minorities was stable; common themes that emerged during the content analysis of a random subsample of 3300 tweets included: racism and blame, anti-racism, and effect on daily life |
| Odlum et al[82] | June | USA | Twitter | 2 558 474 Tweets from Jan 21 to May 3, 2020 | Clustering algorithm; NLP; network diagrams | 15 topics (in four themes) were identified; positive sentiments, cohesively encouraging online discussions, and behaviours for COVID-19 prevention were uniquely observed in African American Twitter communities |
| Park et al[83] | May | South Korea | Twitter | 43 832 unique users and 78 233 relationships on Feb 29, 2020 | Network analysis; content analysis | Spread of information was faster in the COVID-19 network than in the other networks; tweets containing medically framed news articles were more popular than were tweets that included news articles adopting non-medical frames |
| Pastor[84] | April | Philippines | Twitter | Tweets were collected on three Tuesdays in March, 2020, since lockdown in Philippines | NLP for sentiment analysis | Negative sentiments increased over time in lockdown |
| Samuel et al[85] | June | USA | Twitter | 900 000 tweets from February to March, 2020 | Sentiment analysis packages; textual analytics; machine learning classification methods: Naive Bayes and logistic regression | For short tweets, classification accuracy was 91% with Naive Bayes whereas accuracy was 74% with logistic regression; both methods showed weaker performance for longer tweets |
| Samuel et al[86] | August | USA | Twitter | 293 597 tweets, 90 variables | Textual analytics to analyse public sentiment support; sentiment analysis by use of R package Syuzhet (version 1.0.6) | For the reopening of the US economy, there was more positive sentiment support than there was negative support; developed a novel sentiment polarity based public sentiment scenarios framework |
| Su et al[87] | June | China and Italy | Sina Weibo and Twitter | 850 Sina Weibo users with posts published from Jan 9 to Feb 5, 2020; 14 269 tweets from 188 unique Twitter users from Feb 23 to March 21, 2020 | Wilcoxon tests | Individuals focused more on home and expressed a high level of cognitive process after a lockdown in both Wuhan, China, and Lombardy, Italy; level of stress decreased, and the attention to leisure increased in Lombardy, Italy, after the lockdown; attention to group, religion, and emotions became more prevalent in Wuhan, China, after the lockdown |
| Thelwall and Thelwall[88] | May | UK | Twitter | 3 038 026 English tweets from March 10 to March 23, 2020 | Word frequency comparison; $\chi^2$ analysis | Women were more likely to tweet about the virus in the context of family, physical distancing, and health care, whereas men were more likely to tweet about sports cancellations, the global spread of the virus, and political reactions |
| Wang et al[89] | July | China | Sina Weibo | 999 978 randomly selected Sina Weibo posts that were related to COVID-19 from Jan 1 to Feb 18, 2020 | Unsupervised Bidirectional Encoder Representations from Transformers model: classify sentiment categories; Term Frequency-Inverse Document Frequency model: summarise the topics of posts; trend analysis; thematic analysis | People were concerned about four aspects regarding COVID-19: the virus origin, symptoms, production activity, and public health control |
| Wicke and Bolognesi[90] | September | Ireland | Twitter | 203 756 tweets | Topic modelling | Although the family frame covers a wider portion of topics, among the figurative frames, war (a highly conventional one) was the frame used most frequently; yet, this frame does not seem to be appropriate to elaborate the discourse around some aspects that are involved in the situation |

(Table continues on next page)

COVID-19.[99] 12 articles studied infodemics that were related to COVID-19 that were circulating on social media platforms. Rovetta and Bhagavathula[42] analysed over 2 million queries from Google Trends and Instagram between Feb 20 and May 6, 2020. Their findings showed that as global interest for COVID-19 information increased, so did its infodemic.[42] Gallotti and colleagues analysed over 100 million tweets and identified that, even before the onset of the COVID-19 pandemic, infodemics threatened public health, although not to the same extent.[37] Pulido and colleagues sampled and analysed 942 tweets, which revealed that although false information had a higher number of tweets, it also had less retweets and lower engagement than did tweets comprising scientific evidence or factual statements.[41] Kouzy and colleagues[39] investigated the extent to which misinformation or unverifiable information about the COVID-19 pandemic was spread on Twitter by analysing 673 English tweets. Their results showed that misinformation accounted for 24·8% (153 of 617) of all serious tweets (ie, not humour-related posts). Health-care or public health accounts had the lowest amount of misinformation; yet still 12·3% (7 of 57) of their tweets included unverifiable information. Moscadelli and colleagues[40] collected and reviewed 2102 news articles that were circulated on the internet. Their analysis showed that fake news was shared over 2 million times, which accounted for 23·1% (2 352 585 of 10 184 351) of total shares between Dec 31, 2019, and April 30, 2020.[40] Similarly, another quantitative study by Galhardi and colleagues comparing the proportion of fake news shared on WhatsApp, Instagram, and Facebook in Brazil showed that fake news was mainly shared on WhatsApp.[36] A UK study by Ahmed and colleagues analysed 22 785 tweets posted by 11 333 Twitter users with #FilmYourHospital to identify and evaluate the source of the conspiracy theory on Twitter.[32] Their work uncovered that ordinary people were the major driver behind the spread of conspiracy theories.[32] Another study investigated the 5G and COVID-19 conspiracy theory that was circulating on Twitter with a random subsample of 233 tweets. The content analysis showed that 34·8% (81) of tweets linked 5G and COVID-19 and 32·2% (75) condemned such theory.[33] Similar research by Bruns and colleagues investigated 89 664 distinct Facebook posts in Australia that were related to this conspiracy from Jan 1 to April 12, 2020, by use of time series and network analysis.[35] The results showed that this conspiracy went viral after March 19, 2020, with unusual coalition among various groups on Facebook. Islam and colleagues analysed 2311 infodemic reports that were related to COVID-19 from Dec 31, 2019, to April 5, 2020, and showed that misinformation was mainly driven by rumours, stigma, and conspiracy theories that were circulating on various social media and other online platforms.[38] Associations between infodemic and bot activities on social media are another important research direction. One study analysed 12 million tweets from the USA and 15 million tweets from the Philippines from March 5 to March 19, 2020, and both countries showed a positive relation between bot activities and rate of hate speech in communities that are denser and more isolated than others.[43] Brennen and colleagues qualitatively analysed 96 samples of visuals (ie, image or video) from January to March, 2020, and categorised misinformation into six trends, noting that, fortunately, there has been no involvement of artificial intelligence deepfake techniques (ie, techniques used to make synthetic videos that closely resemble real videos) so far.[34]

|  | Publication month | Origin | Social media | Study population and sample size | Methods | Key findings |
| --- | --- | --- | --- | --- | --- | --- |
| (Continued from previous page) | | | | | | |
| Xi et al[91] | September | China | Sina Weibo | 188 unique topics, their views, and comments from Jan 20 to April 28, 2020 | Thematic analysis; temporal analysis | Six themes were identified: the most prominent theme was older people contributing to the community (46 [24%] of 188) followed by older patients (defined by keywords—eg, "older people", "old-aged people", "grandmother", "grandfather", "old grandmother", "old grandfather", "old woman", and "old man") in hospitals (43 [23%]); the theme of contributing to the community was the most dominant in the first phase (Jan 20–Feb 20, 2020; period of COVID-19 outbreak in China); the theme of older patients in hospitals was most dominant in the second (Feb 21–March 17, 2020; turnover period) and third phase (March 18–April 28, 2020; post-peak period in China) |
| Xie et al[92] | August | China | Baidu Search Index and Google Trends | Number of cases by Feb 29, 2020: 79 968 cumulative confirmed cases, 41 675 cured cases, 2873 dead cases | Kendall's $T_b$ rank test | Both the Baidu Search Index and Google Trends indices showed a similar trend in a slightly different way; daily Google Trends were correlated to seven indicators, whereas daily Baidu Search Index was correlated to only three indicators; these indexes and rumours are statistically related to disease-related indicators; information symmetry was also noted |
| Xue et al[93] | November | Canada | Twitter | 1 015 874 tweets from April 12 to July 16, 2020 | LDA | Nine themes about family violence were identified |
| Yigitcanlar et al[94] | October | Australia | Twitter | 96 666 tweets from Australia in Jan 1 to May 4, 2020 | Descriptive analysis; content analysis; sentiment analysis; spatial analysis | Social media analytics is an efficient approach to capture attitudes and perceptions of the public during a pandemic; crowdsourced social media data can guide interventions and decisions of the authorities during a pandemic; effective use of government social media channels can help the public to follow the introduced measures and restrictions |
| Yu et al[95] | July | Spain | Twitter | 22 223 tweets | Topic modelling; network analysis | Identified eight news frames for each newspaper's Twitter account; the entire pandemic development process is divided into three periods: precrisis, lockdown, and recovery period; understanding of how Spanish news media cover public health crises on social media platforms |
| Zhao et al[96] | May | China | Sina Weibo and microblog hot search list | 4056 topics from Dec 31, 2019, to Feb 20, 2020 | Word segmentation; word frequency; sentiment analysis | The trend of public attention could be divided into three stages; the hot topic keywords of public attention at each stage were slightly different; the emotional tendency of the public towards the COVID-19 pandemic-related hot topics changed from negative to neutral between January and February, 2020, with negative emotions weakening and positive emotions increasing overall; COVID-19 topics with the most public concern were divided into five categories: the situation of the new cases of COVID-19 and its effects, front-line reporting of the pandemic and the measures of prevention and control, expert interpretation and discussion on the source of infection, medical services on the front line of the pandemic, and focus on the pandemic and the search for suspected cases |
| Zhu et al[97] | July | China | Sina Weibo | 1 858 288 microblog data | LDA | A so-called double peaks feature appeared in the search curve for epidemic topics; the topic changed over time, the fluctuation of topic discussion rate gradually decreased; political and economic centres attracted high attention on social media; the existence of the subject of rumours enabled people to have more communication and discussion |
| All studies were published in 2020. LDA=latent Dirichlet allocation. NLP=natural language processing. | | | | | | |

*Table*: Summary of chosen articles

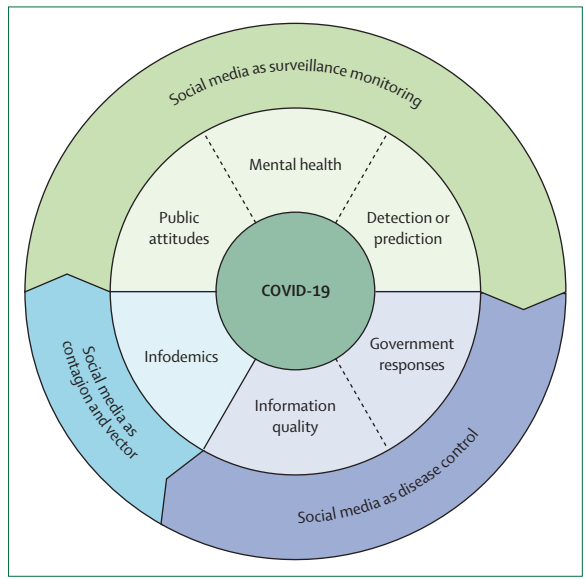

*Figure 2:* Modified Social Media and Public Health Epidemic and Response framework

### Social media for surveillance and monitoring

Three themes emerged under this category: public attitudes, mental health, and detection or prediction of COVID-19 cases. Public attitudes and mental health are reflections regarding the public perceptions and mental health effects of the pandemic; detection or prediction of COVID-19 cases includes typical surveillance studies aiming to propose ways to detect or predict COVID-19 cases.

48 selected articles gauged the attitudes and emotions that were expressed by social media users regarding the COVID-19 pandemic, mainly by use of content and sentiment analysis. Twitter accounted for 33 articles and Sina Weibo accounted for 8 articles. Public attitude can be further divided into the following sub-themes: public sentiment towards the COVID-19 pandemic and interventions, stigma and racism, and ageism.

To learn about the public sentiment towards the overall COVID-19 pandemic and its interventions, Abd-Alrazaq and colleagues[7] analysed 167 073 unique English tweets that were divided into four categories: origin, source, regional and global effects on people and society, and methods to reduce transmission of SARS-CoV-2. Tweets regarding economic loss had the highest mean number of likes, whereas travel bans and warnings had the lowest number of likes.[7] Kwon and colleagues investigated 259 529 English tweets in the USA, using trending and spatiotemporal analyses, and noted that tweets about social disruptiveness had the highest number of retweets, whereas tweets about COVID-19 interventions had the highest number of likes.[73] A content analysis of 522 Reddit comments showed that the topic of symptoms accounted for 27% (141) of all comments, followed by the topic of prevention (25% [131]).[74] Likewise, another content analysis of 155 353 unique English tweets showed that the most mentioned topic was "peril of COVID-19".[76] Additionally, a study that examined 126 049 English tweets by use of sentiment analysis and latent Dirichlet analysis for topic modelling showed that the most common emotion that was mentioned was fear, and the most common topic that was mentioned was the economic and political effects.[79] Al-Rawi and colleagues studied emojis in over 50 million tweets and identified five primary subjects: morbidity fears, health concerns, employment and financial issues, praise for front-line workers, and unique gendered emoji use.[51] Samuel and colleagues investigated 293 597 tweets with sentiment analysis and noted more positive emotions than negative emotions towards the US economy reopening.[86] Analysing 2 558 474 English tweets by use of clustering and network analyses, Odlum and colleagues identified that African Americans shared positive sentiments and encouraged virtual discussions and prevention behaviours.[82] A study investigated gender differences in terms of topics by analysing 3 038 026 English tweets.[88] The results showed that tweets from women were more likely to be about family, physical distancing, and health care, whereas tweets from men were more likely to be about sports cancellations, pandemic severity, and politics. In Canada, Xue and colleagues analysed 1 015 874 tweets via latent Dirichlet analysis to identify nine themes about family violence.[93] In Australia, Yigitcanlar and colleagues analysed 96 666 tweets and identified that the public's attitude could be captured efficiently through social media analytics.[94] One qualitative content analysis of 30 profiles from Instagram, Twitter, and TikTok in Brunei identified five types of attitudes towards physical distancing: fear, responsibility, annoyance, fun, and resistance.[80] In Turkey, to show the effects of social media on human psychology and behaviour, Arpaci and colleagues[52] used evolutionary clustering analysis on 43 million tweets between March 22 and March 30, 2020. The study suggested that high-frequency word clusters, such as death, test, spread, and lockdown denoted the public's underlying fear of infection and death from the virus, whereas terms such as stay home and social distancing corresponded to behavioural shifts.[52] A study in Luzon, Philippines,[84] in which sentiment analysis was done by use of natural language processing, showed that most Filipino Twitter users expressed negative emotions towards COVID-19, and the negative mood grew stronger over time in lockdown.[84] Sentiment analysis of 107 990 English tweets uncovered that a negative feeling towards the COVID-19 pandemic dominated, and topic modelling showed three major themes in people's concerns: the COVID-19 pandemic emergency, how to control COVID-19, and reports on COVID-19.[54] Another study analysed 373 908 Belgian tweets and retweets, which showed that the public relied on the EU coalition to tackle the pandemic.[72] De Santis and colleagues analysed 1 044 645 tweets to identify daily hot topics in Italy that were related to the COVID-19 pandemic and developed a framework for prospective research.[62] One thematic analysis study of 1 920 593 Arabic tweets in Egypt showed that negative emotions and sadness were high in tweets showing affective discussions, and the dominant themes included the outbreak of the pandemic, metaphysics responses, signs and symptoms in confirmed cases, and conspiracism.[64] In Singapore, Lwin and colleagues examined 20 325 929 tweets using sentiment analysis and showed that public emotions shifted over time: from fear to anger and from sadness to gratefulness.[77] Chang and colleagues examined over 1·07 million Chinese texts from various online sources in Taiwan using deductive analysis and identified that negative sentiments mainly came from online news with stigmatising language linked with the COVID-19 pandemic.[56] In India, one study investigated 410 643 tweets via sentiment analysis and latent Dirichlet analysis and showed that positive emotions were overall substantially higher than negative sentiments, but this observation diminished at individual levels.[61] Another study analysed 29 554 tweets from the second lockdown (ie, April 15–May 3, 2020) and 47 672 tweets from the third lockdown (ie, May 4–May 17, 2020) via sentiment analysis uncovered positive attitudes towards the second lockdown but negative attitudes towards the third lockdown in India.[57] One study analysed 868 posts from Reddit and noted sentiments to be 50% (434) neutral, 22% (191) positive, and 28% (243) negative in India.[63] A study in South Korea examined 43 832 unique users and their relations on Twitter by use of content and network analyses and showed that tweets including medical news were more popular than tweets containing non-medical news.[83] A study from Ireland analysed 203 756 tweets through topic modelling and identified that war was the most frequently used frame for the pandemic.[90] In the USA, Damiano and colleagues qualitatively analysed 600 English tweets and showed neutral sentiment across most tweets.[59] Politics also had an essential role in shaping people's opinion.[59] A study of 19 803 tweets from Democrats and 11 084 tweets from Republicans by use of random forest in the USA showed that Democrats put more emphasis on public health and direct aid to US workers, whereas Republicans put more emphasis on national unity, China, and businesses.[67] Results of a study involving various online data sources from Italy, the UK, the USA, and Canada showed that media was the major driver of the public's attention, but attention decreased with saturation of the media with news about COVID-19.[66] Compared with other users, Reddit users focused more on health, data related to new disease, and preventative interventions. Researchers in Spain studied 22 223 tweets by use of topic modelling and network analysis.[95] They identified eight frames and noted that the entire pandemic could be divided into three periods: precrisis, lockdown, and recovery periods. Using 563 079 English Reddit posts that were related to COVID-19, Jelodar and colleagues proposed a novel method to detect meaningful latent topics and sentiment–comment classification.[69] Samuel and colleagues examined over 900 000 tweets to study the accuracy of tweet classifications among logistic regression and Naive Bayes methods.[85] They identified that Naive Bayes had 91% of accuracy compared with 74% from the logistic regression model.[85]

Han and colleagues analysed 1 413 297 Sina Weibo posts and observed that the public paid attention to information regarding the epidemic, especially in metro areas.[68] Zhao and colleagues studied 4056 topics from the Sina Microblog hot search list and noted that the public emotions shifted from negative to neutral to positive over time and that five major public concerns existed: the situation of the new cases of COVID-19 and its effects, front-line reporting of the pandemic and the measures of prevention and control, expert interpretation and discussion on the source of infection, medical services on the front line of the pandemic, and focus on the pandemic and the search for suspected cases.[96] Li and colleagues[75] did an observational infoveillance study with a linear regression model by analysing 115 299 Sina Weibo posts. The results showed that the number of Sina Weibo posts positively correlated with the number of reported cases of COVID-19 in Wuhan. Additionally, the qualitative analysis classified the topics into the following four overarching themes: cause of the virus, epidemiological characteristics of COVID-19, public responses, and others.[75] Chen and colleagues examined relationships between citizen engagement through government social media and media richness, dialogic loop, content type, and emotion valence.[58] Citizen engagement through government social media refers to sum of shares, likes, and comments in this study, so the higher the sum, the greater the citizen engagement through government social media. Media richness quantifies how much information that a sender transfers to a receiver via a medium and is based on the media richness theory (ie, "the potential information load of communication media, emhasising the abilities of promoting shared meaning").[101] Dialogic loop, or dialogic communication theory, is defined as an approach that promotes a dialogue between a speaker and audience. According to the American Psychological Association, emotion valence refers to "the value associated with a stimulus, expressed on a continuum from pleasant to unpleasant or from attractive to aversive".[100] For instance, happiness is typically considered to be pleasant valence. Chen and colleagues analysed 1411 posts that were related to COVID-19 from Healthy China, an official account of the National Health Commission of China on Sina Weibo. Findings showed an inverse association between media richness and citizen engagement through government social media, indicating that posts with plain texts had higher citizen engagement through government social media than did posts with pictures or videos. A positive association between dialogic loop and citizen engagement through government social media was noted, as evidenced by 96% (1355 of 1411) of responses to these posts having hashtags and 25% (353 of 1411) containing questions. In terms of media richness, when posts had both a high media richness and positive emotion, citizen engagement through government social media increased, whereas when posts had a high media richness and negative emotion, citizen engagement decreased. Regarding content type, when posts were related to the latest news about the pandemic, stronger negative emotions led to increased citizen engagement through government social media.[58] Yin and colleagues[65] proposed a new multiple-information susceptible-discussing-immune model to analyse the public opinion propagation of COVID-19 from Sina Weibo posts that were collected from Dec 31, 2019, to Feb 27, 2020. The researchers reported that the reproduction rate of this proposed model reached 1·78 in the early stage of COVID-19 but decreased to around 0·97 and was maintained at this level. Such a result showed that the information on COVID-19 would continue to increase slowly in the future until it stabilises. However, this stability would depend on how much information is received on COVID-19. Wang and colleagues[89] analysed 999 978 randomly selected Sina Weibo posts that were related to COVID-19 through an unsupervised Bidirectional Encoder Representations from Transformers model for sentiments and a term frequency-inverse document frequency model for topic modelling. The authors identified four public concerns: the virus origin, symptom, production activity, and public health control in China.[89] Xi and colleagues examined 241 topics with their views and comments via thematic and temporal analysis and noted that older adults contributing to the community was the most frequent theme in the first phase of COVID-19 in China (ie, Jan 20–Feb 20, 2020).[91] The theme of older patients in hospitals was most frequent in the second (ie, Feb 21–March 17, 2020) and third phase (ie, March 18–April 28, 2020). Using Wilcoxon tests, Su and colleagues examined posts from 850 Sina Weibo users and 14 269 tweets from Italy.[87] The findings showed that Italian people paid more attention to leisure, whereas Chinese people paid more attention to the community, religion, and emotions after lockdowns. Analysing the top 200 accounts from WeChat via regressions and content analysis, Ma and colleagues showed that both non-medical and medical reports had positive effects on people's behaviours.[78] Using Kendall's Tau-B rank test, Xie and colleagues investigated relations among the Baidu Attention Index, daily Google Trends, and numbers of COVID-19 cases and deaths.[92] Daily Google Trends were correlated to seven indicators, whereas daily Baidu Search Index was correlated only to three indicators.[92] Zhu and colleagues analysed 1 858 288 Sina Weibo posts and noted that topics changed over time but political and economic posts attracted greater attention than did other topics.[97]

Regarding stigma and racism, Kim[71] analysed 27 849 individual tweets in South Korea by use of a binary logistic regression to gauge network size and semantic network analysis to capture contextual and subjective factors. The results indicated that size of personal social network was inversely correlated with impolite language use. Namely, users with larger social networks were less likely to post uncivil messages on Twitter than were users with smaller social networks. This study suggested that the size of the social network influenced the language choice of social media users in their postings.[71] Research compared public stigma before and after the introduction of the terms Chinese virus or China virus in 16 535 English tweets from before introduction and 177 327 tweets from after introduction.[55] The results showed an almost 10 times increase, nationwide and statewide and in the USA, from 0·38 tweets posted per 10 000 people referencing the two terms before introduction to 4·08 tweets posted per 10 000 after introduction. A similar study examined 339 063 tweets from non-Asian respondents via local polynomial regression and interrupted time-series analysis.[60] The findings showed that, when stigmatising terms, such as Chinese virus, were used by media (starting from March 8, 2020), the bias index (ie, Implicit Americanness Bias) began to increase, and such bias was more profound in conservatives than in members of any other political subgroup. Nguyen and colleagues analysed 3 377 295 tweets that were related to race in the USA using sentiment analysis and uncovered a 68·4% increase in negative tweets referring to Asian people, whereas tweets referring to other races remained stable.[81]

Regarding ageism, a study[70] investigating Twitter content that was related to both COVID-19 and older adults analysed a random sample of 351 English tweets. 21·1% (74) of the tweets implied diminished regard for older adults by downplaying or dismissing concerns over the high fatality of COVID-19 in this population.[70] Similar research examined 188 tweets via thematic analysis and showed that 90% (169) of tweets opposed ageism, whereas 5% (9) of tweets favoured ageism, and 5% (10) of tweets were neutral.[53]

Two of 81 reviewed studies, both based in China, focused on assessing the mental health of social media users.[44,45] A cross-sectional study[44] investigated the relationship between anxiety and social media exposure, which is theoretically defined as "the extent to which audience members have encountered specific messages".[102] The researchers distributed an online survey based on the Chinese version of WHO-Five Well-Being Index for depression and the Chinese version of Generalized Anxiety Disorder Scale for anxiety. Respondents included 4872 Chinese citizens aged 18 years and older from 31 provinces and autonomous regions in China. After controlling for all covariates through a multivariable logistic regression, the study showed that frequent social media exposure increased the odds ratio of anxiety, showing that frequent social media exposure is potentially contributing to mental health problems during the COVID-19 outbreak.[44] To explore how people's mental health was influenced by COVID-19, Li and colleagues[45] analysed posts from 17 865 active Sina Weibo users to compare sentiments before and after declaration of COVID-19 outbreak by the National Health Commission in China on Jan 20, 2020. The researchers identified increased negative sentiments, including anxiety, depression, and indignation, after the declaration and decreased positive sentiments expressed in the Oxford happiness score. Additionally, cognitive indicators showed increased sensitivity to social risks but decreased life satisfaction after the declaration.[45]

Six of 81 studies investigated the detection or prediction of COVID-19 outbreaks with social media data. Qin and colleagues[22] attempted to predict the number of newly suspected or confirmed COVID-19 cases by collecting social media search indexes for symptoms (eg, dry cough, fever, and chest distress), coronavirus, and pneumonia. The data were analysed by use of subset selection, forward selection, lasso regression, ridge regression, and elastic net. Results showed that the optimal model was constructed via the subset selection. The lagged social media search indexes were a predictor of new suspected COVID-19 cases and could be detected 6–9 days before confirmation of new cases.[22] To evaluate the possibility of early prediction of COVID-19 cases via internet searches and social media data, Li and colleagues[17] used the keywords coronavirus and pneumonia to retrieve corresponding trend data from Google Trends, Baidu Search Index, and Sina Weibo Index. By use of the lag correlation, the results showed that the correlation between trend data with the keyword coronavirus and number of laboratory-confirmed cases was highest 8–12 days before increase in confirmed COVID-19 cases in the three platforms. Similarly, the correlation between trend data for the keyword coronavirus and new suspected COVID-19 cases was highest 6–8 days before increase in new suspected cases. The correlation between trend data for the keyword pneumonia and new suspected cases was highest 8–10 days before increase in new suspected COVID-19 cases across the three platforms.[17] Peng and colleagues studied 1200 Sina Weibo records using spatiotemporal analysis, kernel density analysis, and ordinary least square regression and noted that scattered infection, community spread, and full-scale outbreak were three phases of early COVID-19 transmission in Wuhan, China.[21] Older people are at high risk of severe COVID-19 and accounted for over 50% of help seeking on Sina Weibo. To identify COVID-19 patients with poor prognosis, Liu and colleagues analysed Sina Weibo messages from 599 patients along with telephone follow-ups.[18] The findings suggested risk factors involving older age, diffuse distribution of pneumonia, and hypoxaemia. A regression study analysed Google Trends searches, Wikipedia page views, and tweets and showed that current Wikipedia page views, tweets from a week before, and Google Trends searches from two weeks before can be used to model the number of COVID-19 cases. To model the number of deaths, all three variables should be one week earlier than for cases.[19]

### Social media as disease control
To inoculate the public against misinformation, public health organisations and governments should create and spread accurate information on social media because social media has had an increasingly important role in policy announcement and health education. Six of 81 articles were categorised as government responses because they examined how government messages and health education material were generated and consumed on social media platforms. Two studies analysed data from Sina Weibo,[23,27] and the other four studies analysed data from Twitter.[28–31]

Zhu and colleagues[23] measured the attention of Chinese netizens—ie, citizen of the net—to COVID-19 by analysing 1101 Sina Weibo posts. They noted that Chinese netizens paid little attention to the disease until the Chinese Government acknowledged and declared the COVID-19 outbreak on Jan 20, 2020. Since then, high levels of social media traffic occurred when Wuhan, China, began its quarantine (Jan 23–Jan 24, 2020), during a Red Cross Society of China scandal (Feb 1, 2020), and following the death of Li Wenliang (Feb 6–Feb 7, 2020).[23] Li and colleagues[27] collected 36 746 Sina Weibo posts to identify and categorise the situational information using support vector machines, Naive Bayes, and random forest as well as features in predicting the number of reports using linear regression. Except for posts that were categorised as counter rumours (ie, used to oppose rumours), they identified that the higher the word count, the more reposts there were. Likewise, posts from unverified users had more reposts for all categories than did posts from verified users, excluding the counter rumours. For counter rumours, reposts increased with the number of followers and if the followers were from urban areas.[27] A qualitative content analysis was done to investigate how G7 leaders used Twitter for matters concerning the COVID-19 pandemic by collecting 203 tweets.[29] The findings showed that 166 of 203 tweets were informative, 48 tweets were linked to official government resources, 19 (9·4%) tweets were morale-boosting, and 14 (6·9%) tweets were political.[29] To assess the political partisan polarisation in Canada regarding COVID-19, Merkley and colleagues[28] randomly sampled 1260 tweets from the social media of 292 federal members of parliament and collected 87 Google Trends for the search term coronavirus. 2499 Canadian respondents aged 18 years and above were also surveyed. The results showed that, regardless of party affiliation, members of parliament emphasised the importance of measures for physical distancing and proper hand-hygiene practices to cope with the COVID-19 pandemic, without tweets exaggerating concerns or misinformation about COVID-19. Search interest in COVID-19 among municipalities was strongly determined by socioeconomic and urban factors rather than Conservative Party vote share.[28] Sutton and colleagues studied 149 335 tweets from public health, emergency management, and elected officials and observed that the underlying emotion of messages changed positively and negatively over time.[30] Wang and colleagues investigated 13 598 tweets that were related to COVID-19 via temporal and network analyses.[31] They categorised 16 types of messages and identified inconsistent and incongruent messages expressed in four crucial prevention topics: mask wearing, risk assessments, stay at home order, and disinfectants or sanitisers.

Eight chosen studies investigated the quality (ie, the number of recommended prevention behaviours that were covered in the videos—eg, wearing a facemask, washing hands, physical distancing, etc) of YouTube videos with COVID-19 prevention information. Basch and colleagues[24] did a cross-sectional study and retrieved the top 100 YouTube videos with the most views that were uploaded in January, 2020, with the keyword of coronavirus in English, with English subtitles, or in Spanish. These 100 videos generated over 125 million views in total. However, fewer than 33 videos included any of the seven key prevention behaviours that are recommended by the US Centers for Disease Control and Prevention.[24] A follow-up study with the same criteria and a successive sampling design gathered the top 100 YouTube videos that were most viewed in January and March, 2020.[25] Findings showed that, in total, the January sample generated over 125 million views, and the March sample had over 355 million views. Yet, fewer than 50 videos in either sample contained any of the prevention behaviours that are recommended by the US Centers for Disease Control and Prevention.[25] Additionally, a study investigated the top 100 YouTube videos about do it yourself hand sanitiser with the most views and showed that the average number of daily calls about paediatric poisoning increased substantially in March, 2020, compared with the previous 2 years.[46]

To analyse the information quality on YouTube about the COVID-19 pandemic and to compare the contents in English and Chinese Mandarin videos, Khatri and colleagues[26] collected 150 videos with the keywords 2019 novel coronavirus and Wuhan virus in English and Mandarin. The DISCERN score and the medical information and content index were calculated as a reliable way to measure the quality of health information. The mean DISCERN score for reliability was low: 3·12 of 5·00 for English videos and 3·25 for Mandarin videos. The mean cumulative medical information and content index score of useful videos was also undesirable: 6·71 of 25·00 for English videos and 6·28 for Mandarin videos.[26] In Spain, a similar study of 129 videos in Spanish identified that information in videos about preventing COVID-19 was usually incomplete and differed according to the type of authorship (ie, mass media, health professionals, individual users, and others).[47] Likewise, one study in South Korea noted that misleading videos accounted for 37·14% (39 of 105) of most-viewed videos and had more likes, fewer comments, and longer viewing times than did useful videos.[48] Two studies in Turkey investigated the quality of YouTube videos regarding COVID-19 information in dentistry.[49,50] One of these studies analysed the top 116 English videos with at least 300 views and showed moderate quality and useful information from these videos.[49] The other study, however, showed poor quality for 24 of 55 (43·6%) English videos, whereas good quality accounted for only 2 (3·6%) videos.[50]

### Discussion

Studies on social media data showed our attitudes and mental state to some extent during the COVID-19 crisis. These studies also showed how we generated, consumed, and propagated information on social media platforms when facing the rapid spread of the SARS-CoV-2 and extraordinary measures for the containment. In our Review, public attitudes accounted for nearly 59% (48 of 81) of the reviewed articles. In terms of social media platforms, 56% (45 of 81) of the chosen articles used data from Twitter, followed by Sina Weibo (20% [16 of 81]). Machine learning analyses, such as latent Dirichlet analysis and random forest, were applied in research that studied public attitudes.

We identified six themes on the basis of our modified SPHERE framework, including infodemics, public attitudes, mental health, detection or prediction of COVID-19 cases, government responses to the pandemic, and quality of prevention education videos. However, a common limitation in all chosen studies on social media data is the comparison of data due to differences in quality, such as formats, metrics, or even the definition of common variables (eg, the amount of time required for a post to be on an individuals screen to be counted as a view). For instance, the definition of a view on one social media platform is likely to be different from another. Besides, not every social media platform offers accessible data, like Twitter and Sina Weibo. To address these challenges, the selected studies have controlled for many factors, including social media platforms, languages, locations, time, misspellings, keywords, or hashtags. However, such search strategies resulted in many study limitations, such as non-representative sample sizes, selection bias, cross-sectional study design, or retrospective study design. We also observed that, given the large amount of available data, most studies across all domains sampled small data size for analyses, except for four studies under the theme of public attitudes that analysed over one million posts via machine learning methods. Additionally, data from Twitter and Sina Weibo accounted for over 70% (59 of 81) of our selected studies. Research examining other social media platforms, including Facebook, Instagram, TikTok, Snapchat, and WhatsApp, is scarce due to barriers of data availability and accessibility. We also identified future research topics that are needed for each category during the COVID-19 pandemic. From an infodemics perspective, additional research is needed to investigate how misinformation, rumours, and fake news (eg, anti-mask wearing reports) undermine preventions and compromise public health, although social media companies, such as Twitter and Facebook, have started to remove accounts that are based on misinformation. Bot posts are another topic to be addressed and studies evaluating effective counter-infodemic interventions are also needed.

Articles regarding public attitudes towards the COVID-19 pandemic have shown sentiments that shifted over time. Yet, this theme can be a useful indicator when evaluating interventions, such as physical distancing and wearing masks, that aim to reduce the risk of COVID-19 infection. However, public sentiments had not been incorporated into many intervention studies by the time that we did this Review. When a disease, such as COVID-19, starts spreading and causing negative sentiments, timely, proper, and effective risk communication is needed to help ease people's anxiety or negative attitudes regarding the COVID-19 pandemic, especially through social media.

Mental health is another issue that requires further investigation. Our chosen studies did not address mental health issues on the basis of age, as symptoms and interventions tend to vary with age. Public health measures, such as physical distancing, that were implemented in the COVID-19 pandemic exacerbated risk factors and adverse health behaviours at the individual and population levels. Studies showed that social media data were useful to detect mental health issues at the population level. Due to the early outbreak of COVID-19 and the prevalence of social media use (eg, Sina Weibo and WeChat) in China, two studies reported increased issues of mental health among the Chinese population.[44,45] A similar trend of deteriorating mental health could happen in other regions. At the time of writing, British Columbia has recorded the highest number of overdose deaths in Canada (May, 2020).[103]

In terms of the surveillance of the COVID-19 pandemic, six chosen studies showed methods to detect or predict the number of COVID-19 cases by use of social media data. Accoridng to our Review, unlike other infectious diseases, such as influenza and malaria, COVID-19 has not had real-time monitoring surveillance developed with social media data. It is possible that the pandemic has evolved so rapidly that finding COVID-19 vaccinations or therapies has been prioritised over real-time monitoring surveillance with social media. Besides, scarcity of accurate and reliable data sources might discourage the development of the COVID-19 real-time surveillance. Moreover, whether COVID-19 is a one-time event or will become seasonal, like influenza, is unknown. If COVID-19 becomes seasonal, then it might be meaningful and useful to establish a real-time model to monitor the disease by use of social media data.

Government responses that were distributed via social media have been increasingly crucial in combating infodemics and promoting accurate and reliable information for the public. However, little has been studied about how efficient and effective these official responses are at leading to public belief or behavioural changes. It also remained unknown whether government posts would reach greater numbers of social media users or have greater effects on them than would infodemics.

YouTube has served as one of the major platforms to spread information concerning the control of COVID-19. Nonetheless, our chosen studies showed that most YouTube videos were of undesirable quality because they contained few recommended preventions from governments or public health organisations. The undesirable quality is a worrisome observation if accurate and reliable videos and other types of information are not created and disseminated in a timely manner. Therefore, videos, especially from public health authorities, should include accurate and reliable medical and scientific information and use relevant hashtags to reach a large audience, generate a high number of views, and increase responses. Moreover, our selected studies were limited to YouTube videos only. Additionally, a substantial proportion of the studies were done using Sina Weibo, which, although used by many people, is exclusive to China and might lead to an over-representation of a single country in this Review.

In summary, although our Review has limitations that are embedded from the chosen studies, we recognised six themes that have been studied so far and identified future research directions. Our adopted framework can serve as a fundamental and flexible guideline when studying social media and epidemiology.

### Conclusion

Our Review identified various topics, themes, and methodological approaches in studies on social media and COVID-19. Among the six identified themes, public attitudes comprised most of the articles. Among the selected studies, Twitter was the leading social media platform, followed by Sina Weibo. Few studies included machine learning methods, whereas most studies used traditional statistical methods. Unlike influenza, we were not able to find studies documenting real-time surveillance that was developed with social media data on COVID-19. Our Review also identified studies that were related to COVID-19 on infodemics, mental health, and prediction. For COVID-19, accurate and reliable information through social media platforms can have a crucial role in tackling infodemics, misinformation, and rumours. Additionally, real-time surveillance from social media about COVID-19 can be an important tool in the armamentarium of interventions by public health agencies and organisations.

### Search strategy and selection criteria

The search strategies used index terms, where applicable, and free-text terms to capture two concepts: social media, including both general terms and specific platform names (eg, Twitter, Facebook, Sina Weibo, and YouTube); and COVID-19. For each database, both indexed terms (ie, MeSH and Emtree) and natural language keywords were used with Boolean operators (ie, AND, OR, and NOT) and truncations (panel). Since each database has distinctive search functionality, individually tailored search statements were developed with appropriate search filters for each database. Final search statements, along with a list of search results, were downloaded from each database. Articles were included if they discussed the use of social media for COVID-19 research and if they were original, empirical studies. Only peer-reviewed articles, including peer-reviewed preprints, in English or Mandarin, were included. A decision to include Chinese publications was based on the fact that COVID-19 cases were first reported in Wuhan, China, and many initial and relevant studies were published in Mandarin; therefore, we wanted to capture most studies regarding the use of social media for COVID-19 research. All articles published between Nov 1, 2019, and Nov 4, 2020, were included. Publications such as reviews, opinion pieces, books, book chapters, articles and preprints that were not peer-reviewed, and articles that were written in languages other than English or Mandarin were automatically excluded. The final reference list was generated on the basis of originality and relevance to the broad scope of this review.